**Defect engineering in oxide heterostructures by enhanced oxygen surface exchange**

By *M. Huijben, G. Koster,\* M.K. Kruize, S. Wenderich, J. Verbeeck, S. Bals, E. Slooten, B. Shi, H.J.A. Molegraaf, J.E. Kleibeuker, S. Van Aert, J.B. Goedkoop, A. Brinkman, D.H.A. Blank, M.S. Golden, G. Van Tendeloo, H. Hilgenkamp,* and *G. Rijnders*

[\*]    Dr. M. Huijben, Dr. G. Koster, M.K. Kruize, S. Wenderich, Dr. H.J.A. Molegraaf, Dr. J.E. Kleibeuker, Prof. A. Brinkman, Prof. D.H.A. Blank, Prof. H. Hilgenkamp, Prof. G. Rijnders
Faculty of Science and Technology and MESA+ Institute for Nanotechnology
University of Twente
7500 AE, Enschede, The Netherlands
E-mail: g.koster@utwente.nl
Dr. J. Verbeeck, Dr. S. Bals, Dr. S. van Aert, Prof. G. van Tendeloo
Electron Microscopy for Materials Science (EMAT)
University of Antwerp
2020 Antwerp, Belgium
E. Slooten, B. Shi, Dr. J.B. Goedkoop, Prof. M.S. Golden
Van der Waals-Zeeman Institute
University of Amsterdam
1090 GL, Amsterdam, The Netherlands

The synthesis of materials with well-controlled composition and structure improves our understanding of their intrinsic electrical transport properties. Recent developments in atomically controlled growth have been shown to be crucial in enabling the study of new physical phenomena in epitaxial oxide heterostructures. Nevertheless, these phenomena can be influenced by the presence of defects that act as extrinsic sources of both doping and impurity scattering. Control over the nature and density of such defects is therefore necessary, are we to fully understand the intrinsic materials properties and exploit them in future device technologies. Here, we show that incorporation of a strontium copper oxide nano-layer strongly reduces the impurity scattering at conducting interfaces in oxide $LaAlO_3$-$SrTiO_3$(001) heterostructures, opening the door to high carrier mobility materials. We propose that this remote cuprate layer facilitates enhanced suppression of oxygen defects by reducing the kinetic barrier for oxygen exchange in the hetero-interfacial film system. This design concept of controlled defect engineering can be of significant importance in applications in which enhanced oxygen surface exchange plays a crucial role.

1. Introduction

Advances in material growth have enabled progressive control of the crystalline quality and impurity density of scientifically and technologically important materials. Traditionally, single crystals have been used to study the intrinsic transport properties of complex oxide materials, in which electron-electron correlation effects, electron-lattice interactions and orbital physics are important. Single crystal growth occurs near thermal equilibrium, allowing for a minimal amount of impurities and defects, which is advantageous for transport and other experiments. For example, doped bulk $SrTiO_3$ crystals exhibit Shubnikov-deHaas (SdH) oscillations[1] for sufficiently low dopant concentrations (Nb, La, oxygen vacancies). In recent years such quantum oscillations have also been observed in oxide thin films, for example $SrRuO_3$ [2] and $SrTiO_3$ [3], where in particular molecular beam epitaxy has been successful in reducing the presence of extrinsic scatterers in the system. The next step of dimensional reduction to ultra-thin films, with thicknesses down to the atomic scale, has generally led to the degradation of the transport properties[4]. A perfect model system for the endeavour of combining ultra-thin film thicknesses with high carrier mobilities has been found in the $LaAlO_3$-$SrTiO_3$ oxide heterostructure system[5,6].

The remarkable electronic transport properties that occur at the interface between the band insulators $SrTiO_3$ and $LaAlO_3$ [5,6,7] have been attributed to the avoidance of a so-called polar catastrophe, which would result from the polarity discontinuity across the nonpolar/polar interface between $SrTiO_3$ and $LaAlO_3$. In the idealized case, a build-up of electric potential within the $LaAlO_3$ would trigger the transfer of electrons from the $LaAlO_3$ surface, through the $LaAlO_3$ layer, into the $SrTiO_3$ conduction band for a $LaAlO_3$ film thickness above a threshold value of 3-4 unit cells[8]. Addition of an extra SrTiO$_3$ capping layer changes the situation dramatically, preventing structural and chemical reconstructions at the LaAlO$_3$ surface and results in metallic behaviour below this threshold, even down to a single LaAlO$_3$ unit cell layer[9,10]. However, to date, although the avoidance of the polar catastrophe is thought by many to lie at the heart of the remarkable physics displayed by these systems, the exact balance between doping via the electronic reconstruction versus from vacancy defects[11,12,13,14,15] or cation intermixing[16,17] is still a matter of ongoing debate.

The exact balance between the different possible sources of electrons residing at or near to the SrTiO$_3$/LaAlO$_3$ interface (Figure 1a), results in reported carrier mobilities of only several thousands cm$^2$V$^{-1}$s$^{-1}$ [18,19]. For the development of systems based on interfaces with such quality suitable for devices, enhanced mobilities are required. In this context, one important challenge is to reduce the number of activated carriers and to further enhance the mobility, via reduction of (extrinsic) sources of carrier scattering. Oxygen vacancies - when present near the two-dimensional electron gas (2DEG) where the mobile states are hosted - act as point defects and strong charge carrier scatterers, in addition to cation defects as well as interface scattering. Therefore, a proven design strategy to minimize and sideline these almost ubiquitous defects would represent a significant leap forward.

In (ultra)thin-film heterointerface systems, the incorporation of oxygen during cool down of the film stack after growth is crucial to minimize the possible presence of oxygen vacancies. To ensure maximal surface oxygen exchange, three main processes have to be optimal: the reaction of oxygen at the surface of the as-grown film; the transfer from the surface into the crystal; and lastly the transport within the crystal[20]. In the first step, molecular oxygen is transformed into oxide ions in the outermost layer of the film stack in a reaction involving both electron transfer and ion transfer. The subsequent chemical diffusion of oxide ions into the bulk involves both ionic and electronic species (i.e. oxygen vacancies and holes), so as to maintain global charge neutrality[21].

Previous studies of oxygen exchange kinetics in $SrTiO_3$ (and variations thereof) have shown that this process can be strongly accelerated by addition of a porous Ag or Pt film on top of the $SrTiO_3$ [22]. This is evidently not a particularly viable process step in the case of heteroepitaxial oxide film stacks such as those studied here. An alternative is to add an overlayer of a layered cuprate such as $YBa_2Cu_3O_{7-\delta}$, at whose surface the reaction steps all occur very fast[23], meaning that the cuprate essentially delivers pre-formed oxide ions to the layer below[20] possibly as a consequence of the high energy of the $O_{2p}$-dominated states at the top of the valence band of the cuprate[24]. Given the similarity in lattice constants of cuprate systems and the $SrTiO_3/LaAlO_3$ basis of the heterointerface system, it should be feasible to grow cuprate layers epitaxially on the $LaAlO_3$, followed by a passivating capping layer of $SrTiO_3$.

In this study we report on the conversion of these ideas into a successful strategy to minimize the deleterious effects of scattering related to oxygen defects in the $LaAlO_3$-$SrTiO_3$ heterointerface. By introducing a remote layer containing strontium copper oxide, as schematically depicted in Figure 1b, the thermally activated carriers found in both the free and $SrTiO_3$-capped heterointerface systems are suppressed, due to effective, cuprate-mediated avoidance of defects at the $LaAlO_3$-$SrTiO_3$ interface. The results are significantly suppressed scattering and greatly enhanced carrier mobilities of the 2DEG states.

**2. Results and discussion**

To compare the transport properties of $SrTiO_3$-$SrCuO_2$-$LaAlO_3$-$SrTiO_3$(001) heterostructures with the generally used $(SrTiO_3)$-$LaAlO_3$-$SrTiO_3$(001) interface structures (with or without $SrTiO_3$ capping layers), $SrTiO_3$-$SrCuO_2$-$LaAlO_3$-$SrTiO_3$(001) heterostructures were fabricated by pulsed laser deposition with reflection high-energy electron diffraction (RHEED) control of the growth process (see the methods section for the deposition settings). Figure 2a shows the RHEED analysis during growth of the $SrTiO_3$-$SrCuO_2$-$LaAlO_3$-$SrTiO_3$(001) heterostructures using pulsed laser deposition. The top panels display the RHEED intensity oscillations during growth of each individual layer indicating successful control on the unit cell (u.c.) scale due to the layer-by-layer growth mode. Results are given for the subsequent growth of a 10 u.c. $LaAlO_3$ layer, a 1,2 or 3 u.c. $SrCuO_2$ layer and a 2 u.c. $SrTiO_3$ toplayer. In Figure 2b the corresponding RHEED patterns are shown for the $TiO_2$-terminated $SrTiO_3$ (100) substrate, $LaAlO_3$ layer, $SrCuO_2$ layer and $SrTiO_3$ toplayer (from left to right), showing conservation of surface structure and low surface roughness. After growth, the heterostructures were slowly cooled to room temperature in $6\times10^{-2}$ mbar of oxygen at a rate of 10 $^{o}$C/minute. X-ray photoelectron spectroscopy indicates that Cu in the $SrCuO_2$ layer for all heterostructures is in a valence state of 2+ (although small contributions of $3d^9L$ configurations cannot be ruled out). In case of $SrTiO_3$-$SrCuO_2$-$LaAlO_3$-$SrTiO_3$(001) heterostructures grown on SrO-terminated $SrTiO_3$(100) substrates, the samples were insulating in standard transport measurements. The low level of surface roughness was confirmed by atomic force microscopy (AFM) analysis of the surface of a 2/1/10 $SrTiO_3$-$SrCuO_2$-$LaAlO_3$-$SrTiO_3$(001) heterostructure. Figure 2c shows the topographic image and the roughness analysis, indicating smooth terraces separated by clear, single unit cell height steps similar to the surface of the initial $TiO_2$-terminated $SrTiO_3$ (100) substrate.

Thus, Figure 2 shows we are able to grow $SrTiO_3$/$LaAlO_3$ interface systems incorporating a cuprate oxide ion 'supplier' layer, which, as we will go on to show later, removes oxygen vacancies from the region of the 2DEG. In Figure 3, we turn to the transport characterization of these systems. All conducting heterostructures exhibit metallic transport behavior down to 2K. In panel (a) the temperature dependence of the carrier density is shown, which has been extracted from the Hall coefficient. The uppermost trace (□) shows the behavior for a 10 unit cell LaAlO$_3$ layer, which displays a large number of activated carriers (~$1.8 \times 10^{14}$ cm$^{-2}$ at room temperature), which freeze out at lower temperatures, resulting in $3 \times 10^{13}$ carriers per cm$^2$ at 2K. This carrier freeze out has been observed in previous LaAlO$_3$-SrTiO$_3$(001) interface studies at an energy scale of 6.0 meV [9], which is comparable to observations in SrTiO$_3$ at low La doping[25]. A typical donor-like defect band located a few meV below the conduction band of SrTiO$_3$ has been associated with the loss of mobile carriers at low temperatures[26], schematically indicated in red in the top inset of Figure 3a.

Addition of a 2 unit cell SrTiO$_3$ cap on a 10 unit cell LaAlO$_3$ film (○), reduces the room temperature carrier density to just above $1 \times 10^{14}$ cm$^{-2}$, and further carrier freeze-out on lowering the temperature brings this system to the same low temperature carrier density as the uncapped system. The solid triangles (stars) show carrier data for one (three) unit cells of SrCuO$_2$ on a 10 unit cell LaAlO$_3$ film, with a 2 unit cell SrTiO$_3$ cap. It is evident that the heterointerfaces containing the remote cuprate layer behave quite differently: the activated carriers are no longer present and both samples with a single and three unit cell thick SrCuO$_2$ layer the interface systems a fully temperature independent carrier concentration of ~$1 \times 10^{13}$ cm$^{-2}$ is observed. This means that the cuprate layer has successfully eliminated the defect-related, donor impurity band and the related thermally activated carriers.

The direct consequence of this successful defect engineering is an improvement in carrier mobility, as shown in Figure 3b. The solid symbols show that (2 unit cell) SrTiO$_3$-capped film systems comprising 9 or 10 LaAlO$_3$ unit cells and 1, 2 or 3 SrCuO$_2$ unit cells all show identical behaviour, with a mobility significantly above that of the cuprate-free films for all temperatures below 150K. At low temperatures, the SrCuO$_2$-containing systems reach carrier mobilities up to 5500 cm$^2$V$^{-1}$s$^{-1}$, a factor five greater than the control films without the cuprate layer. A minimum interfacial carrier density of ~$7\times10^{12}$ cm$^{-2}$ was measured for our heterostructures, which can be re-calculated as a minimum volumetric carrier density of ~$7\times10^{18}$ cm$^{-3}$ by taking an upper limit of 10 nm for the thickness of the conducting interfacial region[27]. Therefore, the observed variations in carrier densities are all in a range well above ~$3\times10^{17}$ cm$^{-3}$, where a maximum carrier mobility for reduced bulk STO is reported[12] and, thus our observation of an increase in carrier mobility on decreasing the carrier density is in good agreement with previously reported studies. This underpins the hypothesis that the donor states in the cuprate-less systems are a source of significant scattering for the 2DEG states, and that these are effectively eliminated by the introduction of the cuprate inter layer. Indeed, similar samples have been investigated using high-field magnetotransport measurements[28], and clear 2DEG sub-band structure has been resolved in these data, attesting to the high quality and low scattering the cuprate layer brings about.

The final panel of Figure 3 compares the critical thicknesses of the LaAlO$_3$ layer required to support metallic conductivity at low temperatures. In the 'first generation', uncapped films of LaAlO$_3$ on SrTiO$_3$, the critical thickness is the well-known figure of four unit cells[8]. On capping the LaAlO$_3$ with SrTiO$_3$ ('second generation' systems), the critical thickness sinks to a single unit cell[9,10]. For the 'third generation' of films of LaAlO$_3$ on SrTiO$_3$ with a cuprate interlayer, Figure 3c shows the critical thickness to be 6 unit cells, indicating a great sensitivity of the conducting channel at and near the SrTiO$_3$/LaAlO$_3$ interface to the electrostatic and chemical termination of the film stack located on top of the LaAlO$_3$ layer. The electrical current in transport measurements does not additionally travel a parallel path through the SrCuO$_2$ layer, as demonstrated by the insulating behaviour in the case of SrTiO$_3$-SrCuO$_2$-LaAlO$_3$-SrTiO$_3$(001) heterostructures grown on SrO-terminated SrTiO$_3$(100) substrates (not shown).

Thus far we have explained the design strategy behind the cuprate layer, and the transport data have shown this approach to be successful in boosting the mobility of the carriers in the 2DEG. We now return to the film itself and present data from both transmission electron microscopy and x-ray absorption spectroscopy that hold a surprise in store as regards the structure of the $SrCuO_2$ inter layer, which proved so effective in reducing the scattering of the 2DEG carriers in the $SrTiO_3$/$LaAlO_3$ heterointerface. We have used atomic resolution scanning transmission electron microscopy combined with electron energy loss spectroscopy (STEM-EELS) to investigate the local composition/structure and crystal lattice parameters of the $SrTiO_3$-$SrCuO_2$-$LaAlO_3$-$SrTiO_3$(001) heterostructures. A high-angle annular dark-field scanning transmission electron microscopy (HAADF-STEM) image of the heterostructure is shown in Figure 4a together with the schematic representation of the individual layers. Clear epitaxial ordering can be observed throughout the complete heterostructure, which is free of structural defects. The observed thicknesses of the individual layers (10 u.c. $LaAlO_3$ + 1 u.c. $SrCuO_2$ + 2 u.c. $SrTiO_3$) measured using STEM match perfectly with the unit-cell controlled growth during PLD deposition monitored using RHEED. A quantitative elemental map of the conducting $LaAlO_3$-$SrTiO_3$(001) interface is shown in Figure 4b, and displays data from the La $M_{4,5}$, Ti $L_{2,3}$ and O K edges, which are enhanced by application of a principal components analysis (PCA) to improve the signal to noise ratio in the EELS spectra. The elemental map and the corresponding La, Ti and O line profiles across the $SrTiO_3$-$LaAlO_3$ interface show minimal La diffusion into the $SrTiO_3$ substrate.[29]

The HAADF-STEM images are of sufficient quality to enable the analysis of possible local structural variations, via direct determination of lattice parameter changes for each unit cell of the heterointerface film stack. To do this we use atomic position quantification from the aberration corrected HAADF-STEM images, building upon the approach recently introduced by Van Aert et al.[30,31] for TEM images. This allows position measurements of all atomic columns with a precision of a few picometers without being restricted by the resolution of the microscope. Figure 5a shows a strip out of a HAADF-STEM image of the $SrTiO_3$-$SrCuO_2$-$LaAlO_3$-$SrTiO_3$(001) heterostructure. The corresponding graph plots the values of the (pseudo)cubic lattice parameter c (normal to the interface), calculated from the atomic column positions in the HAADF-STEM image together with their 95% confidence intervals. Using the lattice parameter of the single crystal $SrTiO_3$ (001) substrate (3.905 Å)[32] as a reference, we determine the c-axis lattice parameter of the grown $LaAlO_3$ layer (~3.75 Å) to be in good agreement with previous results from X-ray diffraction of such thin films[7]. Now turning to the cuprate and $SrTiO_3$ capping layers, the data show that for the former the c-axis lattice parameter is essentially the same as that of the $LaAlO_3$ at 3.75 Å. The top $SrTiO_3$ capping layer exhibits a c-axis lattice parameter of ~4 Å. Given stoichiometric transfer of the $SrCuO_2$ target to the cuprate film, two possible Cu-O networks are possible in the sandwich between the $LaAlO_3$ and $SrTiO_3$ cap. Firstly, the 'infinite layer' structure, comprising of a single $CuO_2$ plane (see lower part of Figure 5c) and out-of-plane Sr ions, which in bulk form has a c-axis lattice parameter of 3.4 Å [33]. The second possibility would be an arrangement in which the $CuO_4$-plaquettes lie in the plane of the film normal, and are arranged as a corner sharing chain, as sketched in the upper part of Figure 5c. This would mean - compared to the infinite layer structure - that oxygen atoms are moving to out-of-plane positions, giving rise to an effective Cu-O and Sr-O layering along the c-direction. As a bulk crystal, this Cu-O network can be found in $Sr_2CuO_3$, and translated to the axis system of the film stack, this kind of corner sharing chain system would possess a c-axis lattice parameter of ~3.9 Å [34]. Based on these observation there is strong support for the plaquette arrangement over the infinite layer. Furthermore, in the field of view of the STEM images, there is no detectable change in the average of the lattice parameter parallel to the interface, consistent with a high-quality, pseudomorphically constrained thin film.

Seeing as the ultrathin $SrCuO_2$ layer is too sensitive to the electron beam to be investigated using EELS in the TEM, we have used polarization-dependent X-ray absorption at the Cu-$L_3$ edge to investigate both 2/1/4 and 2/1/6 $SrTiO_3$-$SrCuO_2$-$LaAlO_3$-$SrTiO_3$(001) heterostructures, with the data for a 2/1/4 sample being shown in Figure 5b. The Cu-$L_3$ edge probes Cu2p→Cu3d excitations and, for a divalent system composed of $CuO_4$-plaquettes the only available final state ($2p^53d^{10}$) involves the $Cu3d_{x2-y2}$ orbital, which lies in the plane of the plaquette (see Figure 5c). Therefore, in the context of the issue at hand, the dichroism in the X-ray absorption between an experiment placing the **E**-vector in and perpendicular to the film plane will provide information on the structure of the cuprate layer. For the infinite layer structure, **E**∥film would yield a strong so-called white line ($2p^63d^9$→$2p^53d^{10}$), and for the **E**-vector pointing out of the film plane, almost no intensity would be expected[35]. For the Cu-O chain comprised of edge-sharing plaquettes, bulk $Sr_2CuO_3$ is a good reference material, and again here the Cu-$L_3$ white line is strong for **E** in the plane of the plaquettes and ten times weaker for **E** perpendicular to the plaquettes[36]. For the data shown in Figure 5b, the X-rays are incident at a grazing angle to the film-stack, with either **E**∥film (linear vertical polarization as shown in Figure 5c) or **E** perpendicular to the film (linear horizontal polarization in Figure 5c). As Figure 5b makes very clear, the Cu-$L_3$ white line is only a little stronger for **E**∥film than for **E**⊥film. This rules out a pure infinite layer structure for the $SrCuO_2$, in agreement with the STEM data of Figure 5a. Over the macroscopic area of 0.1 x 2.6 $mm^2$ probed by the grazing incidence XAS experiment - if we assume equal proportions of Cu-O chains running along the two in-plane axes of the film - the data of Figure 5b would be consistent with more than 60% of the $CuO_4$-plaquettes in a 'standing', or chain-like configuration, and only 40% arranged as in the infinite layer structure.

Recent DFT-based calculations[37] have studied the question of the lowest energy structure for such ultrathin cuprate layers in film stacks such as those under investigation here, coming to the conclusion that the SrCuO$_2$ should be present in the form of edge-sharing chain structures, with the thin film normal lying in the plane of the CuO$_4$-plaquettes. This expectation matches well with both the c-axis distances observed in the STEM data (Figure 5a) as well as with the fact that only weak linear dichroism is observed in the polarized X-ray absorption data (Figure 5b). A preliminary conclusion would be to correlate the chain-like structure of the cuprate layer to the enhanced oxygen uptake, which needs more detailed analysis.

To determine the mechanism of the remote strontiumcopperoxide layer to eliminate defects, additional experiments have been performed. An intermediate cool down step to room temperature in between the depositions of LaAlO$_3$ and SrCuO$_2$ was applied. Due to this additional step, thermally activated carriers were observed, similar to those interface structures without a SrCuO$_2$ layer. This observation also excludes intermixing as the possible cause for creation of defects. In order to further investigate the phenomenology of the strontium copper oxide layer in the elimination of defects, we have performed two additional sets of experiments. Firstly, a series of heterostructures have been fabricated under different oxygen partial pressures during the growth of the LaAlO$_3$ layer. Figure 6a displays carrier concentration data showing that over a wide oxygen pressure regime, from $10^{-3}$ to $10^{-6}$ mbar, the incorporation of a SrCuO$_2$ layer results in the removal of the thermally activated carriers and their associated donor defect levels, leaving a constant low carrier density over the entire temperature range (excluding the effects of oxygen defects in the bulk of the substrate, as reported previously for lowest oxygen partial pressures[5,12]). Interestingly, carrier mobilities in excess of 50.000 cm$^2$V$^{-1}$s$^{-1}$ are observed for the 3rd generation 2DEGs grown in the lowest oxygen partial pressures (Figure 6b).

These results demonstrate a marked dependence of the observed carrier mobility on the oxygen pressure during growth of the heterostructures with a SrCuO$_2$ layer. The differences between those samples with and without the cuprate interlayer allow us to make a separation of the effects on the final carrier mobility observed for the complete heterostructure system: the initial oxidation level of the heterostructure/substrate, the disorder induced by bulk diffusion within the single-crystalline substrate and possibly by the growth mode of the deposited material on the substrate surface. In the following, each of these factors is described in detail.

The oxygen pressure during deposition will determine the initial oxidation level of the $LaAlO_3$ thin film and interface during growth. It is known that in oxide materials oxygen vacancies can easily be formed when a low oxygen pressure is chosen in combination with high temperatures. Therefore, immediately after the growth of the $LaAlO_3$, some amount of oxygen vacancies can be present, and that this amount can be considerable, depending on the actual oxygen pressure used during growth. These vacancies could in part be eliminated during cool-down to room temperature. However, transport measurements for samples without the cuprate interlayer generally show the presence of activated carriers (e.g. see Fig. 6), indicating that oxygen vacancies still exist for all interfaces grown in oxygen pressures in the range of $10^{-6} - 10^{-3}$ mbar. These samples have not been fully oxidized due to a limitation in the oxygen surface exchange. The remaining oxygen vacancies - which are charged impurities - act as scattering centers, limiting the mobility of the mobile charge carriers.

The bulk diffusion within the substrate crystal is also determined by the oxygen pressure used. It is known that Sr-diffusion within a $SrTiO_3$ substrate crystal is minimized at low oxygen pressures[38]. Therefore, for oxide growth at low oxygen pressures such as $10^{-6}$ mbar a highly ordered $TiO_2$-terminated surface of the $SrTiO_3$ crystal can be maintained, resulting in a highly ordered $LaAlO_3$-$SrTiO_3$ interface. In contrast, for higher oxygen pressures of $10^{-3}$ mbar some Sr-diffusion to the $SrTiO_3$ surface could lead to a mixed-termination of $TiO_2$ and SrO, resulting in more disorder at the $LaAlO_3$-$SrTiO_3$ interface. The disorder will influence the scattering of the mobile charge carriers, and therefore, lower the carrier mobility for high oxygen pressures of $10^{-3}$ mbar. Additionally, the growth mode of $LaAlO_3$ on a $SrTiO_3$ substrate exhibits a transition from two-dimensional, layer-by-layer growth at low oxygen pressures to island growth at high oxygen pressures[7]. The actual growth mode will influence the amount of disorder, and therefore the crystallinity, of the final $LaAlO_3$ thin film and thus the interfacial region. Local defects in the crystal structure will have a strong influence on the scattering of mobile charge carriers being transported in the 2D-layer at and near to the interface.

In the work presented here, we use a cuprate interlayer to dramatically enhance the oxygen surface exchange, thereby improving the oxidation of the whole system during cool-down and consequently reducing the amount of oxygen vacancies in our samples. That this factor is operative is clear in the elimination of activated carriers (see Fig. 3), which leads to a significant decrease in the number of transport-active charge carriers observed at low temperature from values between $2.0\text{-}2.7 \times 10^{13}$ $cm^{-2}$ to $0.7\text{-}1.4 \times 10^{13}$ $cm^{-2}$ when this cuprate layer is introduced. This effect is clearly present for all cuprate-containing samples grown in oxygen pressures in the range of $10^{-6} - 10^{-3}$ mbar (see Fig. 6a).

At the same time, the addition of the cuprate layer leads to an increase in the carrier mobility, confirming the reduction of scattering by oxygen vacancies. The actual carrier mobility of the whole system depends on the remaining scattering of the mobile charge carriers from disorder at the interface caused by the growth mode and bulk diffusion at different oxygen pressures. The removal of the masking effect on the mobility due to oxygen-vacancy induced scattering achieved by adding the cuprate layer now enables us to zoom in on the next sources of scattering, which are clearly dependent on the oxygen pressure during growth of the $LaAlO_3$-$SrTiO_3$ interface. Growth in the highest oxygen pressure of $10^{-3}$ mbar results in a degree of disorder at the interface and a maximum carrier mobility of about 5500 $cm^2V^{-1}s^{-1}$. The intermediate oxygen pressure of $10^{-5}$ mbar gives rise to less disorder and a carrier mobility up to about 9000 $cm^2V^{-1}s^{-1}$. The lowest oxygen pressure of $10^{-6}$ mbar gives a region at/near the interface with the lowest disorder, which in turn leads to a very high low-temperature carrier mobility of about 50.000 $cm^2V^{-1}s^{-1}$. Thus, this research takes an important step in showing that the cuprate layer is crucial in the enhancement of the surface oxygen exchange so as to eliminate oxygen vacancies. The final carrier mobilities are then limited by remnant disorder, which can be minimized by growing at low oxygen pressure, such that low bulk diffusion and an ideal layer-by-layer growth give a highly ordered interface with low scattering, and therefore, a high carrier mobility.

A second set of additional experiments involved studying the influence of introducing various different cuprate layers besides $SrCuO_2$ such as $CuO$, $YBa_2Cu_3O_{7-\delta}$ and $La_{1.85}Sr_{0.15}CuO_4$ (data not shown). All of these cuprate layers showed the same strong reduction of impurity scattering signaled by the absence of the carrier freeze-out at low temperatures. Introduction of $LaTiO_3$ or $LaNiO_3$ layers instead of the cuprate layer did not have the same effect, demonstrating the particular efficacy of cuprate layers in enhancing the oxygen incorporation, thereby strongly reducing carrier scattering from oxygen vacancies at the interface.

Finally, it is also important to note that the $SrTiO_3$-$SrCuO_2$-$LaAlO_3$-$SrTiO_3$(001) heterostructures showed metallic behaviour down to 60 mK without any signature of the superconductivity which has been observed previously in delta-doped $SrTiO_3$ [39] and at the $LaAlO_3$-$SrTiO_3$ interface[27]. Seeing as the 2DEG we see in action here (the quantum oscillations of which are studied in detail in Ref. 28) is outside of the superconducting regime in the samples discussed here, this makes it clear that a high mobility quasi-2D electron gas and superconductivity can be observed independently in $SrTiO_3$, unlike has been concluded in an earlier report[39].

## 3. Conclusion

In conclusion, the introduction of a remote SrCuO$_2$ layer in the SrTiO$_3$-capped LaAlO$_3$-SrTiO$_3$ interface system strongly enhanced the electron mobility by eliminating the negative influence of defect states. This suppression of oxygen defects in oxide hetero-interfacial film systems enables the fundamental study of previously inaccessible quantum phenomena in complex oxide 2DEGs. Furthermore, this design concept of controlled defect engineering can be of significant importance in applications in which enhanced oxygen surface exchange plays a crucial role.

**4. Experimental**

Atomically smooth TiO$_2$-terminated SrTiO$_3$(100) substrates were prepared by a combined HF- etching/anneal treatment. All substrates had vicinal angles of ~0.1°. Single-crystal LaAlO$_3$ and SrTiO$_3$ targets as well as a stoichiometric SrCuO$_2$ target were ablated at a laser fluence of 1.3 J/cm$^2$ and a repetition rate of 1 Hz. During growth, the substrate was held at 850 °C in an oxygen environment at $2\times10^{-3}$ mbar for LaAlO$_3$ (except when mentioned otherwise, while for SrCuO$_2$ and SrTiO$_3$ the conditions were adjusted to 650 °C and $6\times10^{-2}$ mbar. The sheet carrier density and mobility were determined by a Hall measurement using a Van der Pauw configuration.

Atomic resolution STEM-EELS and HAADF-STEM measurements were performed using the Qu-Ant-EM microscope at the University of Antwerp consisting of a probe-corrected TITAN G2 80-300 (FEI) instrument equipped with a GIF Quantum spectrometer for electron energy loss spectroscopy (EELS). The effective probe-size during acquisition is approximately equal to 1.5 A . Low loss and core-loss spectra are recorded quasi-simultaneously by using the spectrometer in dual EELS mode. The collection and convergence angle are $\alpha = 21$ mrad and $\beta = 25$ mrad, respectively. The energy resolution in STEM-EELS was approximately equal to 1.2 eV.

X-ray absorption measurements were carried out at the Cu-$L_{2,3}$ edges using linearly polarised synchrotron radiation from the UE56/2-PGM-2 undulator beamline at the BESSY-II electron storage ring at the Helmholtz-Zentrum Berlin. The absorption cross-section was monitored using the drain current, and normalized in the pre-edge region. The films studied were transferred from the UHV PLD chamber in Twente to the XAS chamber at the light source in Berlin using a UHV sample transfer chamber, in which the pressure was maintained under $5 \times 10^{-10}$ mbar at all times. Linearly polarized synchrotron radiation (spot-size 900 µm horizontal, 100 µm vertical) impinged on the film at an incidence angle of 20 degrees, with an energy bandwidth of 350 meV. At grazing incidence, selecting p-polarized light puts the E-vector of the x-rays in the plane of the film stack, whereas selecting s-polarized light placed the E-vector close to the normal of the heterointerface film structure.

**Acknowledgements**

This work is supported by the Netherlands Organization for Scientific Research (NWO) through VENI (M.H.), VIDI (A.B., G.R.) and VICI (H.H.) grants and by the Dutch Foundation for Fundamental Research on Matter (FOM) through the InterPhase program. We are grateful for experimental support during the XAS experiments to Christian Schuessler-Langeheine and Christoph Trabant, whose experimental station is funded by the BMBF (05K10PK2). Access to HZB was also supported by the EU (7th FP, no. 226716). The Qu-Ant-EM microscope was partly funded by the Hercules fund from the Flemish Governement. G.V.T. acknowledges funding from the European Research Council, ERC grant N°246791 – COUNTATOMS and J.V. acknowledges funding from the ERC Starting Grant 278510 VORTEX. All authors acknowledge funding by the European Union Council under the 7th Framework Program (FP7) grant nr NMP3-LA-2010-246102 IFOX and the Research Foundation Flanders (FWO, Belgium).

Received: ((will be filled in by the editorial staff))
Revised: ((will be filled in by the editorial staff))
Published online: ((will be filled in by the editorial staff))

**Figures**

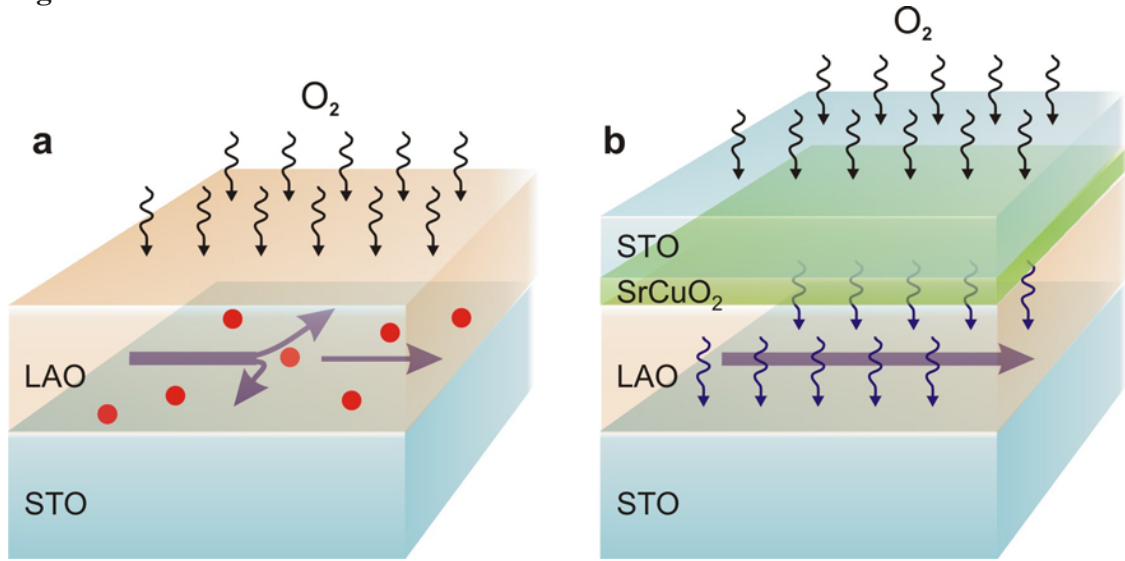

**Figure 1 | Schematic representation of enhanced oxygen incorporation in $SrTiO_3$-$(SrCuO_2)$-$LaAlO_3$-$SrTiO_3$(001) heterostructures.** (a) The limited oxygen surface exchange causes the presence of oxygen vacancies at an $LaAlO_3$-$SrTiO_3$(001) interface leading to defect scattering. (b) The introduction of a $SrCuO_2$ layer enhances the oxygen exchange and eliminates the oxygen vacancies resulting in reduced scattering of the carriers at the interface. Carrier transport at the interface is indicated by blue arrows, while the oxygen vacancies are represented by red points.

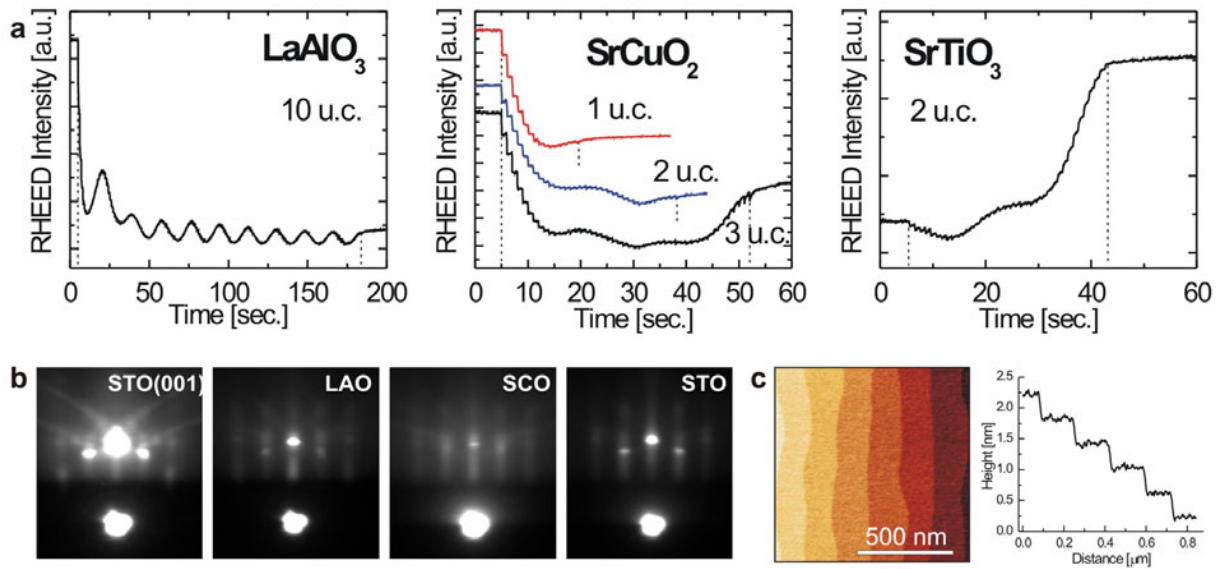

**Figure 2 | Thin film growth of SrTiO$_3$-SrCuO$_2$-LaAlO$_3$-SrTiO$_3$(001) heterostructures by pulsed laser deposition.** (**a**) RHEED intensity monitoring during growth of subsequently a 10 u.c. LaAlO$_3$ layer, a 1,2 or 3 u.c. SrCuO$_2$ layer and a 2 u.c. SrTiO$_3$ toplayer. Clear intensity oscillations indicate layer-by-layer growth of single unit cells. Dashed lines indicate start/stop of laser pulses. (**b**) RHEED patterns after growth of each consecutive layer. (**c**) Surface topography and roughness analysis of a 2/1/10 SrTiO$_3$-SrCuO$_2$-LaAlO$_3$-SrTiO$_3$(001) heterostructure by AFM of a 1x1 μm$^2$ area.

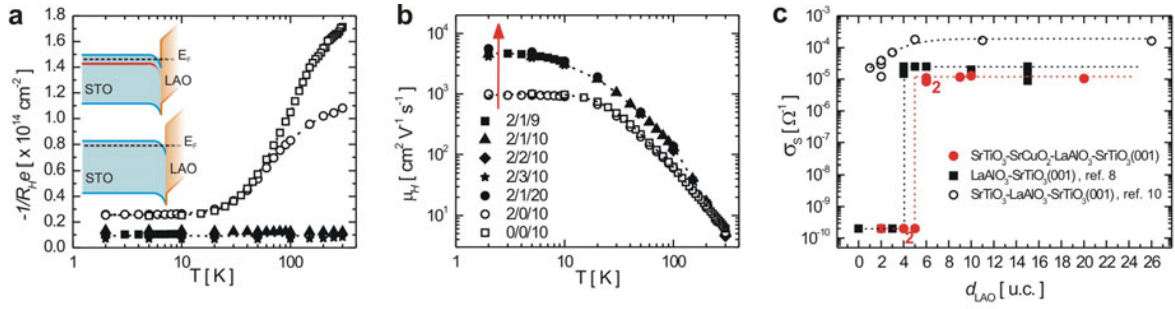

**Figure 3 | Transport properties of SrTiO$_3$-(SrCuO$_2$)-LaAlO$_3$-SrTiO$_3$(001) heterostructures.** (**a**) With or without a SrCuO$_2$ layer. Temperature dependence of $-1/R_H e$, indicating the carrier density, for heterostructures with (closed symbols) and without (open symbols) a SrCuO$_2$ layer. The corresponding bandstructures are schematically represented in the insets. When the additional SrCuO$_2$ layer is introduced, the commonly observed defect donor band (shown in red in the top inset) is eliminated (bottom inset). (**b**) Corresponding temperature dependence of Hall mobility $\mu_H$. Various heterostructure configurations are given, for example 2/1/10 SrTiO$_3$-SrCuO$_2$-LaAlO$_3$-SrTiO$_3$(001) represents a 2 u.c. SrTiO$_3$ toplayer with a 1 u.c. SrCuO$_2$ layer and a 10 u.c. LaAlO$_3$ layer on a SrTiO$_3$ substrate; in the heterostructure indicated with 2/0/10 SrTiO$_3$-SrCuO$_2$-LaAlO$_3$-SrTiO$_3$(001), the SrCuO$_2$ layer is absent. (**c**) Sheet conductance dependence on LaAlO$_3$ layer thickness at 300 K. Heterostructures with a SrCuO$_2$ layer exhibit a sharp insulator-metal transition at a LaAlO$_3$ layer thickness of 6 unit cells. Data for LaAlO$_3$-SrTiO$_3$(001) single interfaces from Thiel *et al.*[8] and for coupled interfaces in SrTiO$_3$-LaAlO$_3$-SrTiO$_3$(001) heterostructures from Pentcheva *et al.*[10] are also shown for comparison.

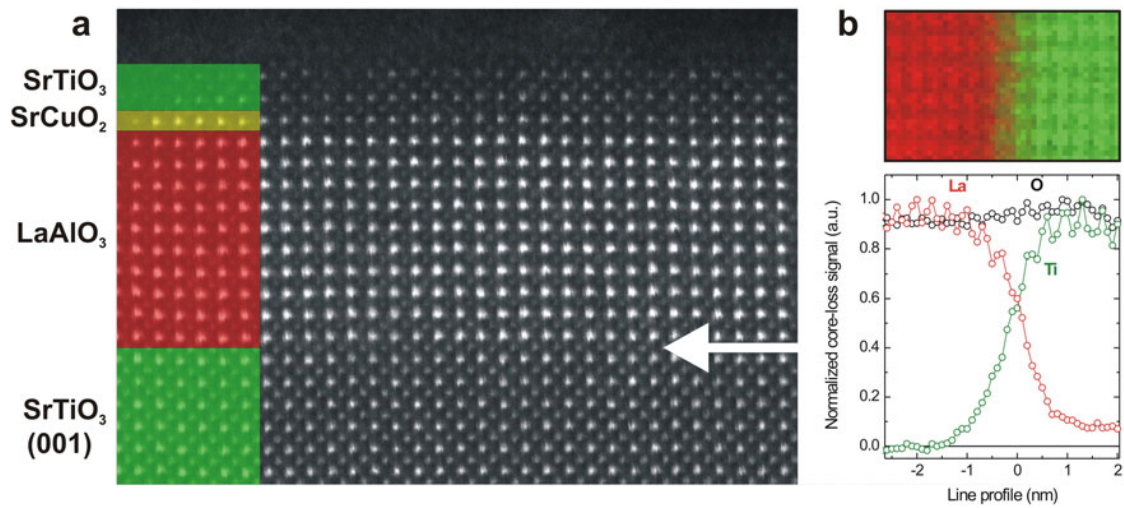

**Figure 4 | Quantitative scanning transmission electron microscopy analysis of the atomic stacking sequences in SrTiO$_3$-SrCuO$_2$-LaAlO$_3$-SrTiO$_3$(001) heterostructures. (a)** HAADF-STEM image of the heterostructure along the [001] zone axis, together with the schematic representation of the individual layers. **(b)** EELS analysis of the conducting LaAlO$_3$-SrTiO$_3$(001) interface within the heterostructure showing normalized core-loss signals for La M$_{4,5}$ (red), Ti L$_{2,3}$ (green) and O K (black) edges.

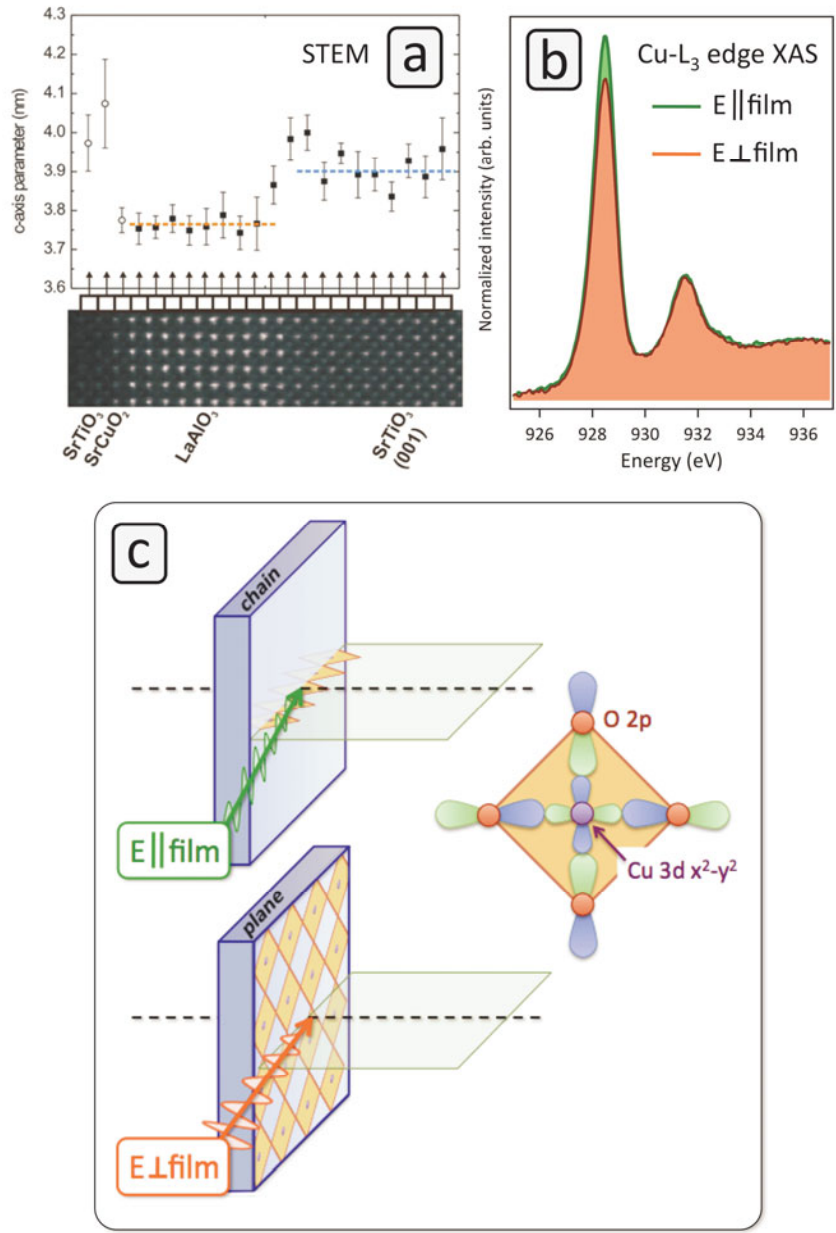

**Figure 5 | Structural ordering of the incorporated SrCuO₂ layer. (a)** Quantitification by STEM of the *c*-axis parameter for the $SrTiO_3$-$SrCuO_2$-$LaAlO_3$-$SrTiO_3$(001) heterostructure using the single crystalline $SrTiO_3$ (001) substrate as a calibration standard. The distances between the atomic planes in $LaAlO_3$ and $SrTiO_3$(001) are indicated as closed squares, while open circles are used for the $SrCuO_2$ and $SrTiO_3$ toplayer. Furthermore, 95% confidence intervals are shown. **(b)** Linear dichroism in X-ray absorption at the Cu-$L_3$ absorption edge for a 2/1/4 $SrTiO_3$-$SrCuO_2$-$LaAlO_3$-$SrTiO_3$(001) heterostructure. The white line at 928.5 eV is only 9% more intense for the polarization vector in the plane of the film, compared to

perpendicular. **(c)** Simplified sketch of possible SrCuO$_2$ structures. Upper: chain-like, such that the E ∥ film case places the E-vector out of the CuO$_4$-plaquettes. Lower: plane-like, such that only the E perpendicular to the film case places the E-vector out of the CuO$_4$-plaquettes. Right: cartoon of the low lying electronic states in a cuprate CuO$_4$ plaquette, showing the Cu3d$_{x^2-y^2}$ and O2p$_{x,y}$ orbitals. These states are the first electron addition states for a divalent cuprate and are polarized in the plaquette.

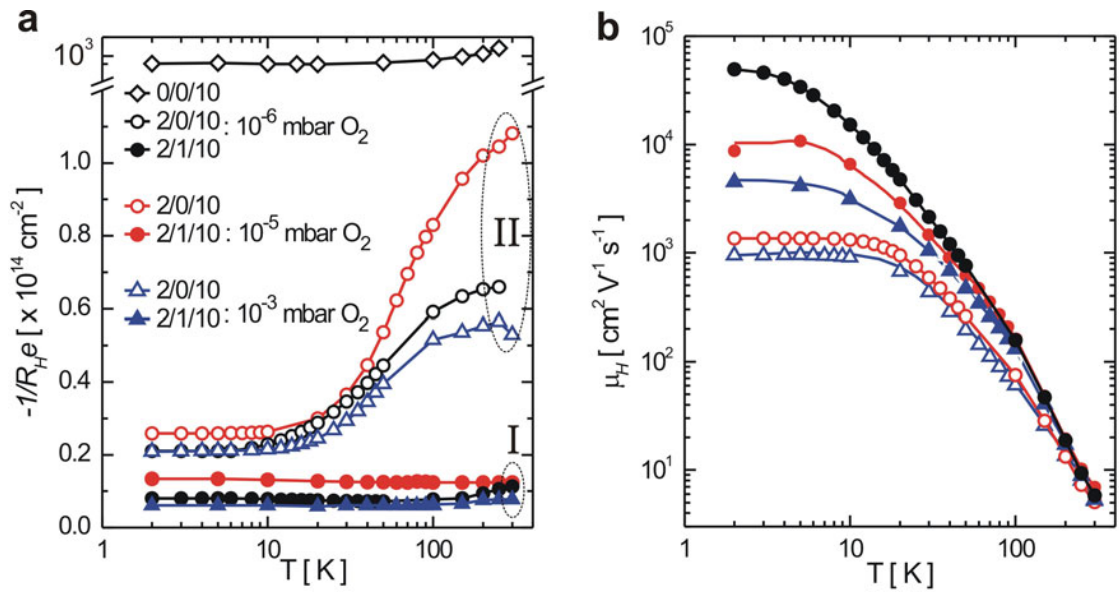

**Figure 6 | Mobility enhancement by defect engineering over a wide oxygen pressure regime.** Temperature dependence of **(a)** $-1/R_H e$, indicating the carrier density, and **(b)** Hall mobility $\mu_H$ for $SrTiO_3$-$(SrCuO_2)$-$LaAlO_3$-$SrTiO_3$(001) heterostructures with (group I: closed symbols) and without (group II: open symbols) a $SrCuO_2$ layer for various oxygen growth pressures. The corresponding Hall resistance versus magnetic field exhibits linear dependence for the complete temperature range (2-300 K), indicating the presence of a single type of carrier. The carrier mobilities of the 2/0/10 and 0/0/10 ($10^{-6}$ mbar $O_2$) samples were excluded as nonlinear dependence is observed in the former and bulk conductivity in the latter.

Submitted to **ADVANCED FUNCTIONAL MATERIALS**

**Info for table of contents**

**Defect engineering in oxide heterostructures by enhanced oxygen surface exchange**

*M. Huijben, G. Koster, M.K. Kruize, S. Wenderich, J. Verbeeck, S. Bals, E. Slooten, B. Shi, H.J.A. Molegraaf, J.E. Kleibeuker, S. Van Aert, J.B. Goedkoop, A. Brinkman, D.H.A. Blank, M.S. Golden, G. Van Tendeloo, H. Hilgenkamp,* and *G. Rijnders*

**Defect engineering of conducting interfaces in oxide $LaAlO_3$-$SrTiO_3$(001) heterostructures by incorporation of a strontium copper oxide nano-layer** strongly reduces the impurity scattering, opening the door to high carrier mobility materials. This remote cuprate layer facilitates enhanced suppression of oxygen defects by reducing the kinetic barrier for oxygen surface exchange in the hetero-interfacial film system.

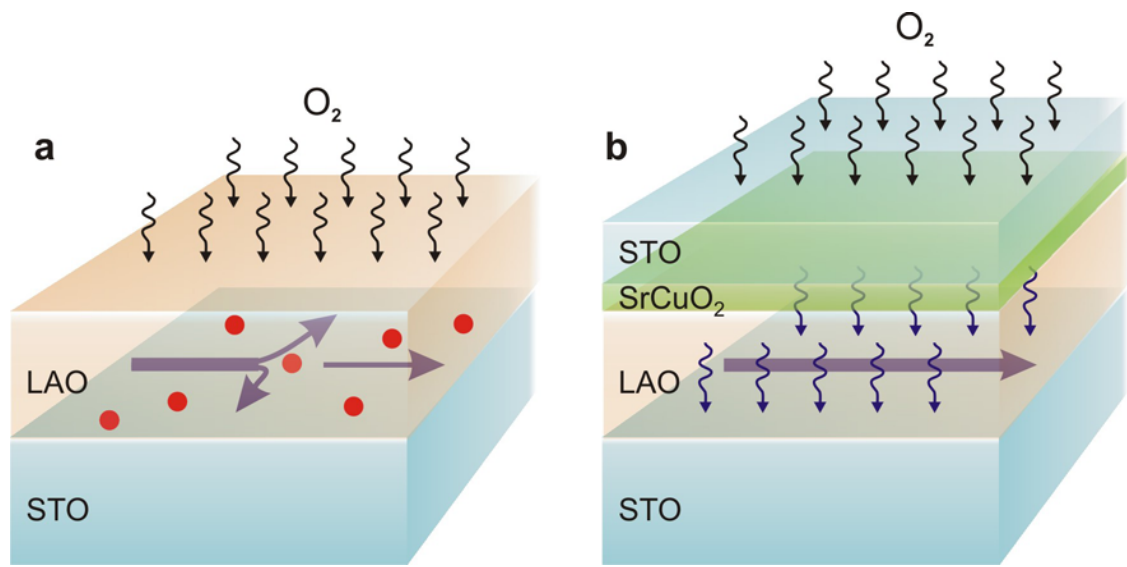